\documentclass[12pt,a4paper]{article}
\usepackage{graphicx}
\usepackage{times}
\title{\bf X-shooter, NACO, and AMBER observations of the LBV Pistol Star \footnote{Based on ESO runs 85.D-0182A, 085.D-0625AC}}
\author{C. Martayan$^{1,2}$, R. Blomme$^3$, J.-B. Le Bouquin$^4$, A. Merand$^1$, G. Montagnier$^1$,\\ 
F. Selman$^1$, J. Girard$^1$, A. Fox$^1$, D. Baade$^5$, Y. Fr\'emat$^3$, A. Lobel$^3$,\\ 
F. Martins$^6$, F. Patru$^1$, T. Rivinius$^1$, H. Sana$^7$, S. Stefl$^1$, J. Zorec$^8$, and T. Semaan$^2$\\
\vspace{1cm}\\
\normalsize $^1$ ESO, Alonso de Cordova 3107 Vitacura, Santiago, Chile\\
\normalsize $^2$ GEPI, Observatoire de Paris, place Jules Janssen 92195 Meudon Cedex, France\\
\normalsize $^3$ Royal Observatory of Belgium, 3 avenue circulaire, 1180 Brussel, Belgium\\
\normalsize $^4$ Laboratoire d’Astrophysique de Grenoble, 38400 Saint-Martin d'Hères, France\\
\normalsize $^5$ ESO, Karl-Schwarschild-Str. 2, Garching bei Muenchen, Germany\\
\normalsize $^6$ GRAAL - UMR5024 Université de Montpellier II  - CC 72 34095 Montpellier Cedex 05 FRANCE\\
\normalsize $^7$ Universiteit van Amsterdam Sterrenkundig 1090 GE Amsterdam The Netherlands\\
\normalsize $^8$ Institut d’Astrophysique de Paris, 98bis boulevard Arago, 75014 Paris, France
}
\date{\mbox{}}
\begin{document}
\maketitle
\pagestyle{empty}
%
%
\def\bull{\vrule height .9ex width .8ex depth -.1ex}
\makeatletter
\def\ps@plain{\let\@mkboth\gobbletwo
\def\@oddhead{}\def\@oddfoot{\hfil\tiny\bull\quad
``The multi-wavelength view of hot, massive stars''; 39$^{\rm th}$ Li\`ege Int.\ Astroph.\ Coll., 12-16 July 2010 \quad\bull}%
\def\@evenhead{}\let\@evenfoot\@oddfoot}
\makeatother
%
%
\def\beginrefer{\section*{References}%
\begin{quotation}\mbox{}\par}
\def\refer#1\par{{\setlength{\parindent}{-\leftmargin}\indent#1\par}}
\def\endrefer{\end{quotation}}
%
%
{\noindent\small{\bf Abstract:} 
We present multi-instruments and multi-wavelengths observations of the famous LBV star Pistol Star. 
These observations are part of a larger program about early O stars at different metallicities.
The Pistol star has been claimed as the most massive star known, with 250 solar masses. 
We present the preliminary results based on X-Shooter 
spectra, as well as the observations performed with the VLTI-AMBER and the VLT-NACO adaptive optics.
The X-shooter spectrograph allows to obtain simultaneously a spectrum from the UV to the K-band with 
a resolving power of $\sim$15000. The preliminary results obtained indicate that Pistol 
Star has similar properties of Eta Car, including shells of matter, but also the binarity.
Other objects of the program, here briefly presented, were selected for their particular nature: 
early O stars with mass discrepancies between stellar evolution models and observations, 
discrepancies with the wind momentum luminosity relation.}
%
%
\section{Introduction}
LBV stars are exceptional transition objects in the stellar evolution of massive stars. Among the emblematic
objects of this category, there is $\eta$ Car. It is now postulated that this kind of star could be a possible aborted
supernova. In all cases, they present a special interest for the stellar evolution theory and models because they
have strong winds, and asymmetries in their structure, are very luminous at the Eddington limit or above.
According to Groh et al. (2009), 2 kinds of ``LBV-phases" could exist: the strong variable LBV
with S-Dor variability that could be very fast rotators rotating close to the breakup velocity and the group of
dormant LBV with less variability like P Cygni. Moreover, the star at the
origin of different SN was a LBV like in SN2006jc (Foley et al. 2007) and SN2006gy (Smith et al. 2007). This direct
evolution from LBV to SN without the step of WR is not yet understood (Smith et al. 2007) but could lead to
the ultra-powerful pair-instability SN. 

\section{Observations}
The observations are spread over several months and periods: 12/2009, 05-06-09/2010.
Early O stars were observed with Xshooter: BRRG56, [ELS2006]N11-026,
[ELS2006]N11-029 in the LMC. These stars were selected as they are expected to be among the earliest O stars in the LMC. 
BRRG56 is classified as an O2 star. They show few absorption lines and several emission lines. In the Galaxy, 
the brightest O star known [BSP2001]8 (Martins et al. 2008) and one of the most massive claimed with 250 M$_{\odot}$ the LBV Pistol 
Star were observed with Xshooter.
In addition, Pistol Star was partially observed with the VLTI-AMBER with the UTs and with the VLT NACO-AO assisted imager. 

\subsection{Brief instruments description}

The VLT-Xshooter (D'Odorico et al. 2006):
It is the first 2nd generation instrument and is a very efficient echelle spectrograph.
One of its interests is the large simultaneous wavelength coverage from the near UV to the K-band with three different arms.
The UVB arm allows the observation of the range $\lambda\lambda$ 290 to 600nm, 
the VIS arm from $\lambda\lambda$ 535 to 1050nm, and the NIR arm from $\lambda\lambda$ 980 to 2500nm.
The resolving power of each arm is defined by the slits used, in the UVB arm from 3300 to 9100, in the VIS arm from 5400 to 19000, 
in the NIR arm from 3500 to 11500.

The VLT-NACO (Lenzen et al. 2003. Rousset et al. 2003) is an adaptive optics assisted imaging in the 1-5 microns. 
It allows the scan of spatial scales from several arcseconds down to 20-15mas. 
The observations were performed with the K, Lp, and NB4 filters.

The VLTI-AMBER (Petrov et al. 2007) is a near-infrared, multi-beam interferometric instrument, 
combining simultaneously up to 3 telescopes.
It allows to scan the spatial scales from 20 to 3 mas.
The observations of Pistol Star were challenging because this star is the faintest object observed at the VLTI and at 
the limit of AMBER capabilities and the fringe tracker FINITO cannot be used.

These 3 instruments were used in order to scan different spatial scales for probing the environment of Pistol Star, 
the nature of Pistol Star itself, and to obtain some parameters/properties of the surrounding nebula.  

\subsection{VLT-Xshooter spectroscopy for the Pistol Star}

\begin{figure}[h!]
\centering
\includegraphics[width=11cm]{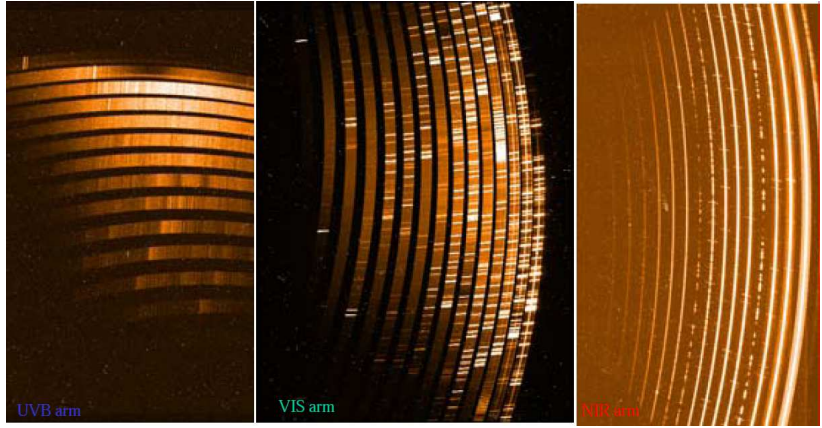}
\caption{Snapshots of the 3 XSHOOTER arms of Pistol Star observations. Left: UVB arm (290-600nm),
middle: VIS arm (525-1050nm), right: NIR arm (980-2500nm). One can see the spectra of Pistol Star mainly in the NIR arm as well as
in the reddest orders of the VIS arm. another very faint object can be seen in the UVB arm.\label{fig1}}
\end{figure}

\begin{figure}[h!]
\centering
\includegraphics[width=6cm,height=15cm, angle=-90]{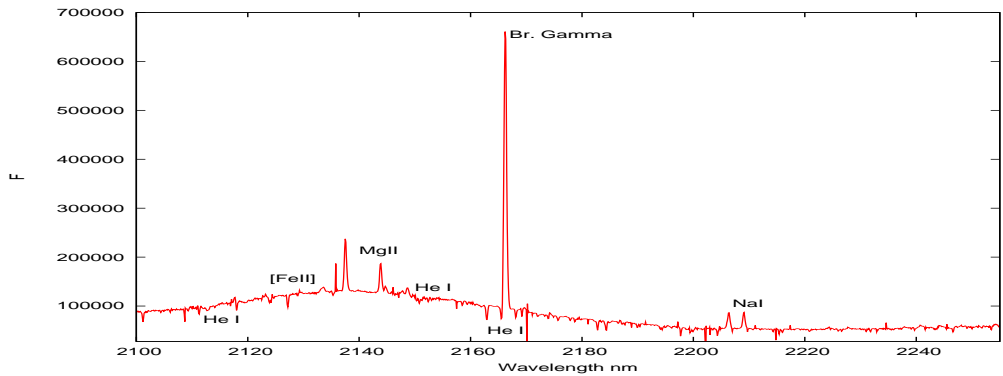}
\caption{Small extract of the XSHOOTER NIR spectrum of Pistol Star in the region close to the Br$\gamma$ line. \label{fig2}}
\end{figure}

The observations in the UVB arm (R used=4000, see Fig. \ref{fig1}-left) show a very faint object of magnitude $\sim$27 in the last 2 red orders (540-600nm). 
The exposure time was 3000s. this object is at about 3" of the Pistol Star. 
It could correspond to a faint star in the NACO images at 2.8"-W.
In the VIS arm (R used=6700, see Fig. \ref{fig1}-middle), Pistol Star appears in the last 5 red orders (770- 1050nm) and disappear in 
the blue orders due to the huge nebulosity in its vicinity, the blue light is absorbed. The exposure time was the same than 
in the UVB arm but here the faint “UVB object” is not seen. The reason is that the UVB arm is more sensitive than the VIS 
one by more than 1 magnitude. There is also another object that can be seen at the edge of the slit (at the right edge of the orders). 
One can also note the large number of skylines present. Up to now no bluer spectrum than 1000nm was obtained 
of Pistol Star.
Finally, in the NIR arm (R=11500), the spectrum displayed in Fig. \ref{fig1}-right was corrected of the sky lines, the exposure time is 50s. 
About 100 NIR exposures were taken. The Pistol Star spectrum corresponds to the central spectrum, there are some emission 
lines present due to the nebulosity surrounding the central star. 
A small extract of the NIR spectrum close to the Br$\gamma$ line is shown in  Fig. \ref{fig2}.
Some emission-lines around Br$\gamma$ can be seen, among them there are [FeII], FeII, MgII, NaI, HeI lines. 
These lines will be compared to the observations by Figer et al. (1999). As a preliminary result, the |EW| of the Br$\gamma$ line is
higher than the measures provided by Figer et al. (1999).
NLTE modelling with CMFGEN (Hillier \& Miller 1998) and other codes for determining the parameters of the star(s), the wind will 
be done. The reduction of the data and the identification of the lines are in progress.

\subsection{Pistol Star and the surrounding stars with the adaptive optics NACO.}
\begin{figure}[h!]
\centering
\includegraphics[width=11cm]{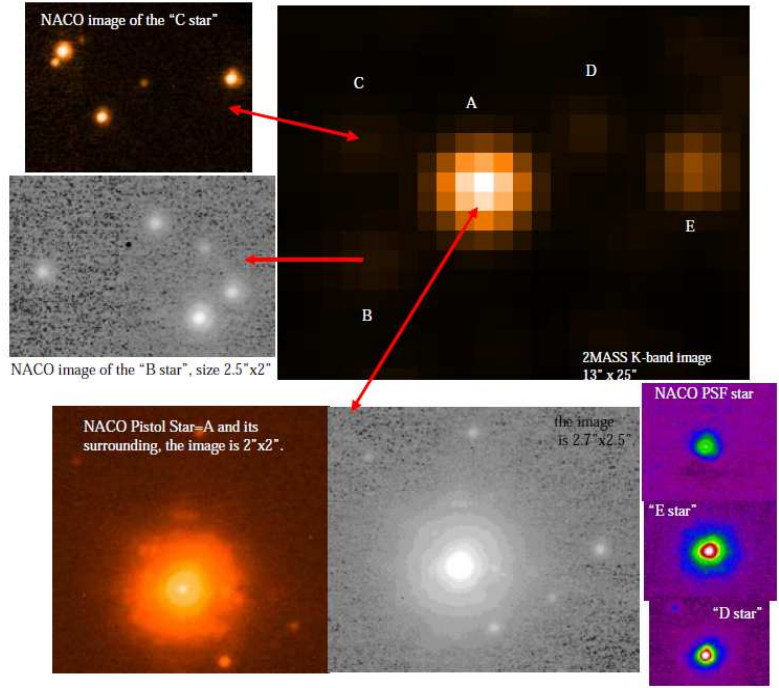}
\caption{NACO observations of Pistol Star and the stars in its vicinity. The top panel shows the 2MASS image and extracts 
of the NACO K band image.
Bottom panel shows Pistol Star with inner small shells of matter in K-band. 
The right column shows different stars including a PSF reference star.  \label{fig3}}
\end{figure}

The NACO images in Fig. \ref{fig3} clearly show that the 2 “B” and “C” 2MASS stars are 2 groups of stars within few arcseconds. 
The “E” star corresponds to the Mira V4644 Sgr and seems to be surrounded by a shell. 
The “D” star corresponds to the WR102e. The “A” star corresponds to Pistol star. 
The star in the right top of the bottom part of Fig. \ref{fig3} shows the NACO PSF. 
One can see several stars close to the central system of Pistol Star at less than 0.4”. Moreover it seems that Pistol 
star is surrounded by different concentric shells of matter. The largest one roughly measures 0.07pc (Pistol Star is at 7.7kpc).
This confirms the spectroscopic result by Figer et al. (1999). 
They found that the spectroscopic lines show the ringlike signature of spherical expansion.

\subsection{Possible binarity of Pistol Star (VLTI-AMBER)}
\begin{figure}[h!]
\centering
\includegraphics[width=8cm]{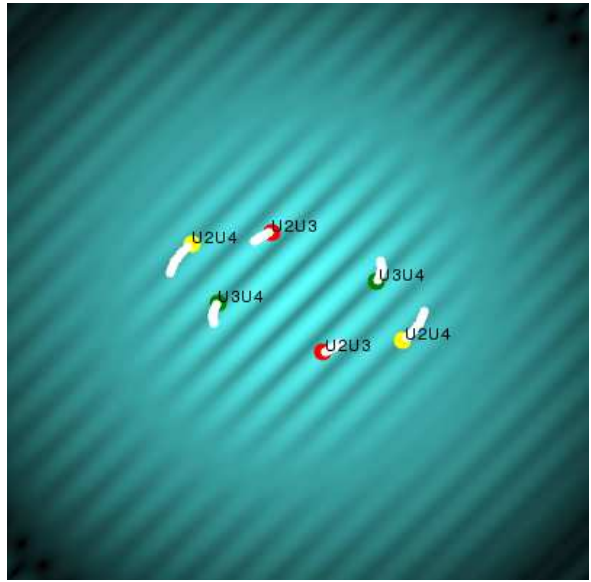}
\caption{UV plane of the Pistol Star observations with the triplet of UTs UT234. There is a companion in the NE at 10-50 mas.\label{fig4}}
\end{figure}

The figure \ref{fig4} shows a rough estimates of the UV plane of the VLTI-AMBER observations. 
There is a departure from single- star morphology. A companion to 
Pistol Star has been detected with a rough separation of 10-50 mas (~80 AU). The reduction and 
the analyse is in progress but this result indicates that like in the Eta Car case (Weigelt et al. 2007),
 the evolution and the understanding of the Pistol Star structure must go through a binary channel. 
Long term monitoring of the system must be done in order to confirm the presence of the companion and 
determine its orbit (check the gravitational link) and the masses of the stars.

\section{Conclusion}
Combining multi-wavelength and multi-technique observations, one can better probe the nature of the objects and better
understand their properties. In the case of Pistol Star, it was found that it has a spectroscopic variability, that
there are small inner shells of matter, and that the star is a binary.
These properties combined to others make of Pistol Star a twin of the LBV $\eta$ Car.
More details will be published in a forthcoming article.

%
%
%
%
\footnotesize
\beginrefer
\refer D'Odorico, S., Dekker, H., Mazzoleni, R., et al. 2006, SPIE, 6269, 626933

\refer Figer, D., Morris, M., Geballe, T. et al. 1999, ApJ, 525, 759

\refer Foley, R., Smith, N., Ganeshalingam, M.,  et al. 2007, ApJ, 657, 105

\refer Groh, J., Damineli, A., Hillier, D., et al. 2009, AJ, 705, L25

\refer Hillier \& Miller 1998, ApJ, 496, 407

\refer Lenzen, R., Hartung, M., Brandner, W., et al. 2003, SPIE 4841, 944

\refer Martins, F., Hillier, D., Paumard, T., et al. 2008, A\&A, 478, 219

\refer Petrov, R., Malbet, F., Weigelt, G., et al. 2007, A\&A, 464, 1

\refer Rousset, G., Lacombe, F., Puget, P., et al. 2003, SPIE 4839, 140

\refer Smith, N., Li, W., Foley, R., et al. 2007, ApJ, 666, 1116

\refer Weigelt, G., Kraus, S., Driebe, T., et al. 2007, A\&A, 464, 87

\endrefer           
\end{document}